\documentclass[11pt]{amsart}
\usepackage{geometry}                
\usepackage{graphicx}
\usepackage{amssymb}
\usepackage{epstopdf}
\usepackage{url}
\usepackage{amsaddr}
\title{The careless use of language in quantum information}
\author{K. Wiesner}
\address{School of Mathematics\\University of Bristol, Bristol, BS8 1TW, U.K.}
 \email{k.wiesner@bristol.ac.uk}
\date{\today}                                           

\begin{document}
\maketitle

An imperative aspect of modern science is that scientific institutions act for the benefit of a common scientific enterprise, rather than for the personal gain of individuals within them. This implies that science should not perpetuate existing or historical unequal social orders. 
%
 Some scientific terminology, though, gives a very different impression. I will give two examples of terminology invented recently for the field of quantum information which  use language associated with subordination, slavery, and racial segregation. 
 
My first example 
 is the term `ancilla qubit'. In a quantum computational algorithm the relevant information is stored in quantum states which, in analogy to `bits' in classical computation, are called `qubits'. Any temporary information to support intermediate work steps   is stored in so-called `ancilla qubits', or simply `ancillas'. 
   `Ancilla' is Latin for `female servant'. A female servant in ancient Rome was a subordinate and very often a slave. The word is not gender neutral and it refers to an unequal social order.  
 The first paper to my knowledge which uses this terminology is an arXiv paper from 1995 \cite{chuang1995quantum}. The term `ancilla' is now standard terminology in the science of quantum information.

My second example of the use of language in quantum information is `quantum supremacy'. It is the name of a subfield in quantum information which has just begun to emerge. This subfield is concerned with the search for tools to computationally simulate quantum systems that are too hard to be simulated with classical computational tools. The hope is to gain insights into the behaviour of highly correlated quantum matter beyond what can be achieved with classical computers. 
The English word `supremacy' denotes the quality or state of having more power, authority, or status than anyone or anything else. These days the word  is closely associated with the politics of `white supremacy' in the apartheid regime of South Africa between 1948 and 1991, known for its institutionalised racial segregation and discrimination. 
The term `quantum supremacy' was coined in 2012 \cite{preskill2012quantum}. It overtly refers to a value system of unequal human rights. Scientific articles and workshops using this term are increasing quickly in number.

Achieving scalable quantum information processing and communication is  considered a strategic goal in Europe \cite{zoller2005quantum} and elsewhere. Within the last 30 years quantum information has  progressed from a mere idea to a very active scientific field drawing upon theoretical and experimental physics, mathematics, computer science, and engineering. It was arguably  Richard Feynman who in 1982 first formulated the potential of quantum mechanics in supporting entirely new and powerful modes of information processing \cite{feynman1982simulating}. 
The field made a leap forward when in 1994 Peter Shor discovered an efficient quantum algorithm for finding the prime factors of large integers, a problem for which no efficient classical algorithm is known\footnote{This result was originally published in proceedings and is now available as \cite{shor1999polynomial}.}. This was a milestone since fast factorisation of integers could be used to break the current public-key cryptography schemes. These schemes are essential for secure bank transfers, credit card payments, or secure data storage.  

So let's halt for a moment and ask ourselves whether using language associated with slavery, misogyny, and racial segregation is really appropriate for a science in the 21st century. 
In a blog post from 2012\footnote{\url{https://quantumfrontiers.com/2012/07/22/supremacy-now/#more-444}}, John Preskill explains his choice of the term `quantum supremacy' and asks fellow scientists to suggest alternatives: 
``In a recent talk, I proposed using the term 'quantum supremacy' to describe super-classical tasks performed using controllable quantum systems. I am not completely happy with this term, and would be glad if readers could suggest something better.'' 
John Preskill's call for an alternative term was not taken up, only the terminology was\footnote{However, see this blog from April 2017: \url{http://dabacon.org/pontiff/?p=11863}.}. 

The association with slavery, misogyny, and racial segregation in these examples is certainly not intended but it is careless. 
%
 Language is powerful. The choice of terminology in science is no exception. We as a scientific community have to think about what we want to stand for and how this is reflected in the way we communicate.

\bibliography{LoQI}{}
\bibliographystyle{plain}

\end{document}